# PERFORMANCE ANALYSIS OF FOURIER AND VECTOR MATRIX MULTIPLY METHODS FOR PHASE RECONSTRUCTION FROM SLOPE MEASUREMENTS

M B Roopashree[1], Akondi Vyas[1,2], B R Prasad[1]
[1]Indian Institute of Astrophysics, 2nd Block, Koramangala, Bangalore
[2]Indian Institute of Science, Bangalore
roopashree@iiap.res.in, vyas@iiap.res.in, brp@iiap.res.in

**Abstract:** The accuracy of wavefront reconstruction from discrete slope measurements depends on the sampling geometry, coherence length of the incoming wavefronts, wavefront sensor specifications and the accuracy of the reconstruction algorithm. Monte Carlo simulations were performed and a comparison of Fourier and Vector Matrix Multiply reconstruction methods was made with respect to these experimental and computational parameters. It was observed that although Fourier reconstruction gave consistent accuracy when coherence length of wavefronts is larger than the corresponding pitch on the wavefront sensor, VMM method gives even better accuracy when the coherence length closely matches with the wavefront sensor pitch.

## 1. INTRODUCTION

Adaptive optics (AO) is used to compensate turbulence induced wavefront distortions in real time. This is achieved by measuring the turbulence induced distortions using a wavefront sensor and applying a cancellation effect using a correcting element usually a deformable mirror [1]. The most commonly used wavefront sensor is the Shack Hartmann Sensor (SHS) which is a two dimensional array of lenslets that measures the local wavefront gradient across each lenslet. A control algorithm converts these local slope measurements into command values that can be addressed to the correcting element. This step called wavefront reconstruction can be considered as the heart of an AO system since it controls the accuracy of wavefront sensing and hence the ability of correcting distortions [2].

The relation between wavefront sensor output and wavefront phase values across the sensor gives rise to a linear system of equations. This can be solved by using standard Vector Matrix Multiplier (VMM) methods [3]. The computation cost of this method increases as $n^2$, where 'n' represents the number of degrees of freedom of the deformable mirror. Very high values of 'n' used in Extreme Adaptive Optics (ExAO) and Multi Conjugate Adaptive Optics (MCAO) makes this method inappropriate for wavefront reconstruction [4]. Fourier method uses the Fourier transform of slope measurements and calculates the wavefront shape in the spatial frequency domain. Then an inverse Fourier transform is used to get back the shape of the wavefront in spatial domain. The computation cost of this method increases as 'n log n' [5].

Codes for the implementation of these two methods were developed and the wavefront reconstruction accuracy as a function of wavefront sensor resolution was studied. This study helps in optimizing the wavefront sensor resolution for a given coherence length of the wavefront. Understanding the sensitivity of wavefront reconstruction accuracy on the sensor resolution guides us in the selection of a suitable method for situations involving fluctuating wavefront coherence length [6]. In this paper, a statistical comparison of these algorithms is presented. Wavefront sampling has a significant effect on wavefront reconstruction accuracy which is described in section 2. The subsequent sections briefly review the VMM and Fourier approaches and present the results.

## 2. SAMPLING GEOMETRY

Local gradients are measured at discrete locations of the wavefront. Depending on the position at which phase differences are estimated with reference to the location of slope measurements, there exist three main classifications in wavefront sampling namely, Fried, Hudgin and Southwell (or Shack Hartmann) geometries extensively discussed in [7-9]. These configurations that illustrate the relationship between positions of slope measurements (in x and y directions) and reconstructed phase are shown in Fig. 1.

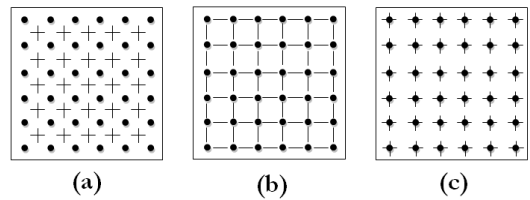

Figure 1. Sampling Geometries (a) Fried (b) Hudgin (c) Southwell

In Fig. 1, horizontal and vertical lines represent positions of slope measurements in x and y directions respectively and dots represent positions of phase estimation. In Fried and Southwell configurations, both x and y slope measurements are made at the same point. In the Shack Hartmann configuration, slope measurement points coincide with the phase evaluation points whereas in Fried configuration, phase evaluation grid is displaced by half the sensor pitch with respect to slope measurement grid. In Hudgin geometry, x and y slope measurement points are displaced from one another and phase is evaluated at the edges of the slopes. The slope to phase relation for different configurations are as follows:

For Fried configuration,

$$S_{ij}^{x} = \frac{[(\Phi_{i+1,j} + \Phi_{i+1,j+1})/2 - (\Phi_{i,j} + \Phi_{i,j+1})/2]}{h}$$

$$S_{ij}^{y} = \frac{[(\Phi_{i,j+1} + \Phi_{i+1,j+1})/2 - (\Phi_{i,j} + \Phi_{i+1,j})/2]}{h}$$

where $i = 1, 2, \ldots, N-1$ and $j = 1, 2, \ldots, N-1$.
(1)

For Hudgin geometry,
$$S_{ij}^{x} = (\Phi_{i+1,j} - \Phi_{i,j})/h$$
where $i = 1, 2, \ldots, N-1$ and $j = 1, 2, \ldots, N$.
$$S_{ij}^{y} = (\Phi_{i,j+1} - \Phi_{i,j})/h$$
where $i = 1, 2, \ldots, N$ and $j = 1, 2, \ldots, N-1$.
(2)

For Shack Hartmann geometry
$$\frac{S_{i+1,j}^{x} + S_{i,j}^{x}}{2} = \frac{\Phi_{i+1,j} - \Phi_{i,j}}{h}$$
where $i = 1, 2, \ldots, N-1$ and $j = 1, 2, \ldots, N$.
$$\frac{S_{i,j+1}^{y} + S_{i,j}^{y}}{2} = \frac{\Phi_{i,j+1} - \Phi_{i,j}}{h}$$
where $i = 1, 2, \ldots, N$ and $j = 1, 2, \ldots, N-1$.
(3)

$S^x$ and $S^y$ represent x and y slope vectors respectively measured at discrete points $(i, j)$. $\Phi_{ij}$ represents phase at $i^{th}$ row and $j^{th}$ column of phase estimation grid. 'N' is the grid dimension which also represents the number of lenslets along a row or column in the SHS (for a square grid) i.e. its resolution. 'h' is the spacing between two nearest points on the phase grid.

## 3. VECTOR MATRIX MULTIPLIER METHOD

Equations (1)-(3) can be represented in matrix formalism as over determined linear system of equations as follows,
$$A \Phi = D S \qquad (4)$$
where 'A' is the co-efficient matrix associated with the phase difference vector. This matrix changes with the sampling geometry. The dimension of the matrix, 'A' is $2N^2 \times N^2$. $\Phi$ is the vector consisting of phase values ($N^2$ values). D is an identity matrix in case of Fried and Hudgin configurations and an averaging matrix in the case of Shack Hartmann configuration. Equation (4) can be solved for '$\Phi$' by computing the generalized inverse of 'A' and pre-multiplying the right hand side of the equation by it.
$$\Phi = (A^{\dagger}A)^{-1}A^{\dagger}DS \qquad (5)$$
$(A^{\dagger}A)^{-1}$ is pre-computed and used in a real time matrix multiplication loop to estimate the phase function. These computations involve operations of the order of $N^4$ (number of degrees of freedom, n = $N^2$).

## 4. FOURIER METHOD

Consider Hudgin geometry for the formulation of Fourier method of wavefront reconstruction. Applying a Discrete Fourier Transform (DFT) on the phase to slope relation in equation (2) and applying the shift property of Fourier transforms yields,

$$\ddot{S}^{x}[k, l] = \ddot{\Phi}[k, l] \left[ \exp\left(\frac{j2\pi k}{N}\right) - 1 \right]$$

$$\ddot{S}^{y}[k, l] = \ddot{\Phi}[k, l] \left[ \exp\left(\frac{j2\pi l}{N}\right) - 1 \right]$$

(6)

where, $\ddot{S}^x$, $\ddot{S}^y$ and $\ddot{\Phi}$ represent the Fourier transforms of $S^x$, $S^y$ and $\Phi$. Spatial frequency coordinates are represented by $(k, l)$. Multiplying $\ddot{S}^x[k, l]$ by $\exp(-j2\pi k/N)$ and $\ddot{S}^y[k, l]$ by $\exp(-j2\pi l/N)$ in eq. (6) and combining them to solve for $\ddot{\Phi}[k, l]$ gives an estimate $\ddot{\tilde{\Phi}}[k, l]$,

$$\ddot{\tilde{\Phi}}[k,l] = \begin{cases} 0; & \text{when } k, l = 0 \\ \left\{ \left[ \exp\left(\frac{-j2\pi k}{N}\right) - 1 \right] \ddot{S}^x[k,l] \right. \\ \left. + \left[ \exp\left(\frac{-j2\pi l}{N}\right) - 1 \right] \ddot{S}^y[k,l] \right\} \\ \times \left[ 4\left( \left(\sin\frac{\pi k}{N}\right)^2 + \left(\sin\frac{\pi l}{N}\right)^2 \right) \right]^{-1}, & \text{else} \end{cases}$$

(7)

The inverse Fourier transform of the expression for $\ddot{\tilde{\Phi}}[k, l]$ in eq. (7) gives the estimate of '$\Phi$'. Similar approach is applied to other configurations also. In particular, for Fried configuration the phase in the Fourier domain is given by,

$\tilde{\Phi}[k, l]$

$$= \begin{cases} 0, & \text{when } k, l = 0, k, l = N/2 \\ \left\{ \left[ \exp\left(\frac{-j2\pi k}{N}\right) - 1 \right] \left[ \exp\left(\frac{j2\pi l}{N}\right) + 1 \right] \ddot{S}^x[k,l] \right. \\ \left. + \left[ \exp\left(\frac{-j2\pi l}{N}\right) - 1 \right] \left[ \exp\left(\frac{j2\pi k}{N}\right) + 1 \right] \ddot{S}^y[k,l] \right\} \\ \times \left[ 2\left( \left(\sin\frac{\pi k}{N}\right)^2 \left(\cos\frac{\pi l}{N}\right)^2 + \left(\sin\frac{\pi l}{N}\right)^2 \left(\cos\frac{\pi k}{N}\right)^2 \right) \right]^{-1}, & \text{else} \end{cases}$$

(8)

The computational cost for the implementation of this method is of the order of $2N^2 \log N$.

## 5. COMPUTATIONAL EXPERIMENTS
### 5.1 Simulations

Monte Carlo simulations were performed to test the accuracy of VMM and Fourier reconstruction at different experimental and computational parameters. These simulations involved four major steps. Firstly, a random phase screen (projection of wavefront on a two dimensional coordinate space) of a given coherence length was simulated. Phase screens were simulated by generating pseudo random numbers following a Gaussian distribution. The number of pseudo random numbers called phase randomness was used as an index for coherence length. Phase randomness of 'M×M' or 'M' represents $M^2$ pseudo random numbers arranged in a two dimensional matrix of size M×M. Greater the phase randomness, lesser the coherence length. As a second step of Monte Carlo simulations, the effect of the phase screen was simulated as shift in the spots of a SHS with known number of subapertures. The wavefront shape was then reconstructed using VMM and Fourier techniques in unison with one of the sampling geometries. Finally, the correlation between the reconstructed wavefront phase and the initially generated random phase screen was computed.

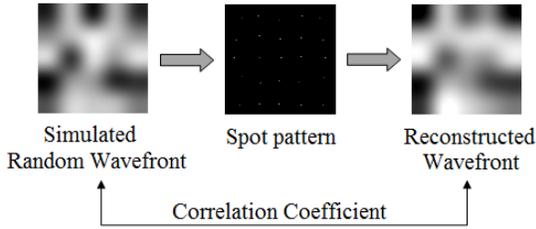

Fig. 2. Steps involved in Monte Carlo simulations are illustrated. Shown simulated random wavefront and reconstructed wavefront are correlated by ~91%

The correlation coefficient indicates the reconstruction accuracy of the wavefronts, higher the correlation, better the reconstruction. The steps involved in the simulations are illustrated in Fig. 2.

### 5.2 Results

For a phase randomness of 12×12, VMM algorithm was applied on all the three configurations by changing the SHS resolution from 2×2 to 30×30. It can be noted from Fig. 3 that VMM method performs best only when the sensor resolution matches with the phase randomness in the phase screen. Southwell geometry which samples like a SHS outperforms other configurations in this case. The consistency of the method in Fried geometry is poor and this is the reason for its worst performance. This can be improved by eliminating associated boundary problems. This behavior can be compared with the Fourier reconstructor as shown in Fig. 4. In the case of Fourier reconstruction, any sensor having a resolution beyond the phase randomness can perform consistently with high correlation coefficient. Maximum correlation for Fourier reconstructor was obtained in the case of Hudgin sampling geometry.

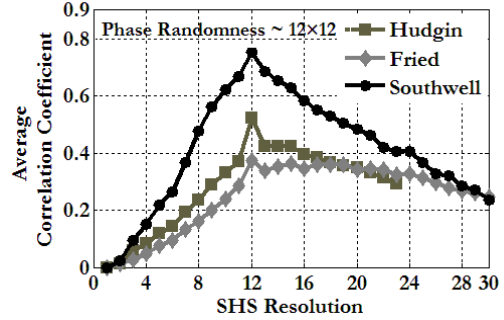

Fig. 3. Performance of VMM reconstructor in different configurations at a phase randomness of 12×12

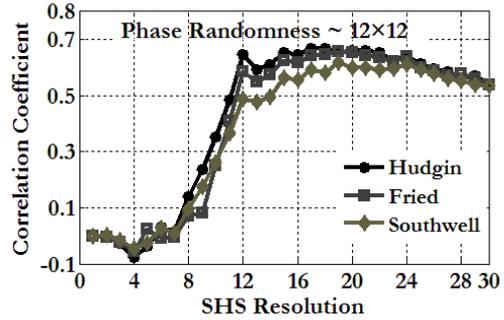

Fig. 4. Performance of Fourier reconstructor in different configurations at a phase randomness of 12×12

The performance of the Fourier reconstructor with changing phase randomness is illustrated in Fig. 5. In this figure, a SHS with resolution 20×20 was used. It can be easily seen that starting from a phase randomness of 10 to 20, the reconstruction accuracy maintained a high value. A similar study was performed in the case of different resolutions of SHS and a similar trend was observed.

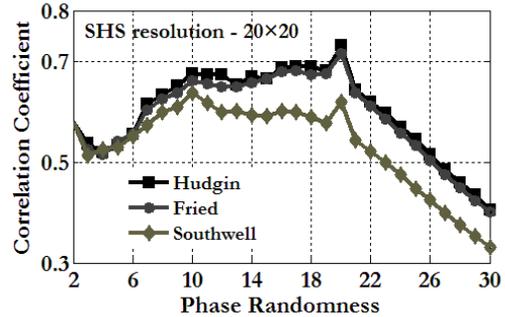

Fig. 5. Performance of Fourier Reconstructor using a SHS of resolution 20×20

Generalizing this observation to a sampling criterion, it can be concluded that Fourier reconstructor performs consistently well when a SHS of resolution N×N is used and the phase maintains a randomness which is between N/2 × N/2 to N×N.

It is appropriate to compare the best performing configurations in both the algorithms. A comparison of VMM (Southwell) and Fourier (Hudgin) methods using 12×12 phase randomness is shown in Fig. 6. The performance of VMM method is better than Fourier reconstructor when the phase randomness exactly matches with the sensor resolution. The advantage of the Fourier method is its consistency of performance for any phase randomness smaller than sensor resolution. In the case of atmospheric adaptive optics, sensors are chosen according to the atmospheric seeing conditions and especially using the prior coherence length measurements. Since the coherence length of atmospheric wavefronts is subject to frequent fluctuations, Fourier method suits well in the case of fluctuating coherence length cases. To improve the reconstruction accuracy and allowing the possibility of using VMM method for greater accuracy, real time measurement of turbulence parameters becomes essential.

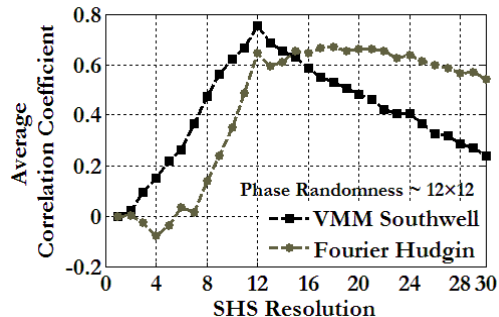

Fig. 6. Comparison of VMM (Southwell) and Fourier (Hudgin) methods for 12×12 phase randomness

### 4. CONCLUSIONS

VMM reconstructor gave best reconstruction accuracy (~85%) when Southwell geometry was used along with least square fitting method to solve the linear system of equations. The performance of reconstruction was relatively poor (~75%) when Fried and Hudgin configuration were used in the case of VMM reconstructor. In general, VMM reconstructor was best when the coherence length matched with the sensor resolution. In contrast, Fourier reconstructor is a little more flexible. Although there is a small peaking when the coherence length matches with the sensor resolution, for a coherence length of $x$, the performance is consistently good for a sensor resolution starting from $x$ to $2x$ irrespective of the sampling geometry. Fourier reconstructor performs best when higher sensor resolution is used. Southwell (~79%) and Hudgin (~78%) geometries suit well in Fourier case.